%
\let\useblackboard=\iftrue
%
%
\newfam\black
\input harvmac.tex
%
\ifx\epsfbox\UnDeFiNeD\message{(NO epsf.tex, FIGURES WILL BE
IGNORED)}
\def\figin#1{\vskip2in}
\else\message{(FIGURES WILL BE INCLUDED)}\def\figin#1{#1}\fi
\def\ifig#1#2#3{\xdef#1{fig.~\the\figno}
\midinsert{\centerline{\figin{#3}}%
\smallskip\centerline{\vbox{\baselineskip12pt
\advance\hsize by -1truein\noindent{\bf Fig.~\the\figno:} #2}}
\bigskip}\endinsert\global\advance\figno by1}
\noblackbox
\def\Title#1#2{\rightline{#1}
\ifx\answ\bigans\nopagenumbers\pageno0\vskip1in%
\baselineskip 15pt plus 1pt minus 1pt
\else
\def\listrefs{\footatend\vskip
1in\immediate\closeout\rfile\writestoppt
\baselineskip=14pt\centerline{{\bf
References}}\bigskip{\frenchspacing%
\parindent=20pt\escapechar=` \input
refs.tmp\vfill\eject}\nonfrenchspacing}
\pageno1\vskip.8in\fi \centerline{\titlefont #2}\vskip .5in}
 
scaled\magstep3
 
scaled\magstep3
 
scaled\magstep3
 
scaled\magstep3
 
scaled\magstep3
\ifx\answ\bigans\def\tcbreak#1{}\else\def\tcbreak#1{\cr&{#1}}\fi

\useblackboard
\message{If you do not have msbm (blackboard bold) fonts,}
\message{change the option at the top of the tex file.}

\font\blackboard=msbm10 scaled \magstep1
\font\blackboards=msbm7
\font\blackboardss=msbm5
\textfont\black=\blackboard
\scriptfont\black=\blackboards
\scriptscriptfont\black=\blackboardss

\else

\fi
%

%
\def\yboxit#1#2{\vbox{\hrule height #1 \hbox{\vrule width #1
\vbox{#2}\vrule width #1 }\hrule height #1 }}
\def\fillbox#1{\hbox to #1{\vbox to #1{\vfil}\hfil}}
\def\ybox{{\lower 1.3pt \yboxit{0.4pt}{\fillbox{8pt}}\hskip-0.2pt}}

\def\comments#1{}

\def\half{{1\over 2}}

\def\Im{{\rm Im\hskip0.1em}}

\def\vev#1{\langle{#1}\rangle}

\def\CN{{\cal N}}

\def\II{\relax{I\kern-.07em I}}

\def\inbar{\,\vrule height1.5ex width.4pt depth0pt}
\def\IZ{\relax\ifmmode\mathchoice
{\hbox{\cmss Z\kern-.4em Z}}{\hbox{\cmss Z\kern-.4em Z}}
{\lower.9pt\hbox{\cmsss Z\kern-.4em Z}}
{\lower1.2pt\hbox{\cmsss Z\kern-.4em Z}}\else{\cmss Z\kern-.4em
Z}\fi}
\def\IB{\relax{\rm I\kern-.18em B}}
\def\IC{{\relax\hbox{$\inbar\kern-.3em{\rm C}$}}}
\def\ID{\relax{\rm I\kern-.18em D}}
\def\IE{\relax{\rm I\kern-.18em E}}
\def\IF{\relax{\rm I\kern-.18em F}}
\def\IG{\relax\hbox{$\inbar\kern-.3em{\rm G}$}}
\def\IGa{\relax\hbox{${\rm I}\kern-.18em\Gamma$}}
\def\IH{\relax{\rm I\kern-.18em H}}
\def\IK{\relax{\rm I\kern-.18em K}}
\def\IP{\relax{\rm I\kern-.18em P}}
\def\pp{{\relax{=\kern-.42em |\kern+.2em}}}

\font\cmss=cmss10 \font\cmsss=cmss10 at 7pt
\def\IR{\relax{\rm I\kern-.18em R}}

\def\Im{{\rm Im\ }}

\def\frac#1#2{{{#1} \over {#2}}}

%
%

\def\NP{{\it Nucl. Phys.\ }}

\def\PL{{\it Phys. Lett.\ }}
\def\PR{{\it Phys. Rev.\ }}

\def\MPL{{\it Mod. Phys. Lett.\ }}

\def\ATMP{{\it ATMP \ }}

\Title{ \vbox{\baselineskip12pt\hbox{hep-th/9811017}
\hbox{BROWN-HET-1150}
}}
{\vbox{
\centerline{Constraints on Higher Derivative Operators}
\centerline{in Maximally Supersymmetric Gauge Theory}}}

\centerline{ David A. Lowe}
\medskip

\centerline{Department of Physics}
\centerline{Brown University}
\centerline{Providence, RI 02912, USA}
\centerline{\tt lowe@het.brown.edu}
\medskip
\centerline{and}
\medskip
\centerline{Rikard von Unge}
\medskip

\centerline{Department of Theoretical Physics and Astrophysics}
\centerline{Masaryk University}
\centerline{Kotl\'{a}\v{r}sk\'{a} 2, 611 37 Brno}
\centerline{Czech Republic}
\centerline{\tt unge@physics.muni.cz}
\bigskip

\centerline{\bf{Abstract}}

\noindent
Following the work of Dine and Seiberg for $SU(2)$, we
study the
leading irrelevant operators on the moduli space 
of $\CN=4$ supersymmetric $SU(N)$ gauge
theory. These operators are argued to be one-loop exact, and are explicitly
computed.

\vfill
\Date{\vbox{\hbox{\sl October, 1998}}}

\lref\dine{M. Dine and N. Seiberg, ``Comments on Higher Derivative
Operators in Some SUSY Field Theories,'' \PL {\bf B409} (1997) 239, hep-th/9705057.}
\lref\mans{M. Henningson, ``Extended superspace, higher derivatives and 
$SL(2,Z)$ duality,'' \NP {\bf B458} (1996) 445, hep-th/9507135.}
\lref\dewit{B. de Wit, M.T. Grisaru and M. Rocek, ``Nonholomorphic
corrections to the one-loop $\CN=2$ super Yang-Mills action,''
\PL {\bf B374} (1996) 297, hep-th/9601115.}
\lref\gonzalez{F. Gonzalez-Rey and M. Rocek, ``Nonholomorphic $\CN=2$
terms in $\CN=4$ SYM: 1-loop calculation in $\CN=2$ superspace,'' \PL
{\bf B434} (1998) 303,
hep-th/9804010.}
\lref\wati{W. Taylor, ``Lectures on D-branes, Gauge Theory and
M(atrices),'' hep-th/9801182.}
\lref\malda{J. Maldacena, ``The Large N Limit of Superconformal Field
Theories and Supergravity,'' \ATMP {\bf 2} (1998) 231, hep-th/9711200.}
\lref\pssone{S. Paban, S. Sethi and M. Stern, ``Constraints From
Extended Supersymmetry in Quantum Mechanics,'' hep-th/9805018.}
\lref\pssthree{S. Paban, S. Sethi and M. Stern, ``Summing Up Instantons
in Three-Dimensional Yang-Mills Theories,'' hep-th/9808119.}
\lref\lowe{D.A. Lowe, ``Constraints on Higher Derivative Operators 
in the Matrix Theory Effective Lagrangian,'' hep-th/9810075.}
\lref\seiberg{N. Seiberg, ``Naturalness versus supersymmetric
non-renormalization theorems,'' hep-ph/9309335, \PL {\bf B318} (1993)
469.}
\lref\buch{E.I Buchbinder, I.L. Buchbinder and
S.M. Kuzenko,``Non-holomorphic effective potential in $N=4$ $SU(n)$
SYM,'' hep-th/9810239.}
\lref\kulik{F. Gonzalez-Rey, B. Kulik, I.Y. Park and M. Rocek,
``Self-Dual Effective Action of $N=4$ Super-Yang Mills,''
hep-th/9810152, v2.}
\lref\butch{I.L. Buchbinder and S.M. Kuzenko, ``Comments on the
background field method in harmonic superspace: non-holomorphic
corrections in $N=4$ SYM,'' \MPL {\bf A13} (1998) 1623, hep-th/9804168.}
\lref\periwal{V. Periwal and R. von Unge, \PL {\bf B430} (1998) 71, hep-th/9801121.}
\lref\diego{D. Bellisai, F. Fucito, M. Matone and G. Travaglini, 
``Non-holomorphic terms in N=2 SUSY Wilsonian actions and
RG equation'' \PR {\bf D56} (1997) 5218, hep-th/9706099.}
\lref\dorey{N. Dorey, V.V. Khoze, M.P. Mattis, J. Slater and W.A. Weir,
``Instantons, Higher-Derivative Terms, and Nonrenormalization Theorems
in Supersymmetric Gauge Theories'', \PL {\bf B408} (1997) 213,
hep-th/9706007.} 

\newsec{Introduction}

The generic actions of field theories with supersymmetry are 
tightly constrained. In certain cases, supersymmetry 
completely determines the form of some of
the terms in the action.
For example with $\CN=2$ in four dimensions
supersymmetry fixes the entire action at leading order at low energies
in terms of a $\CN=2$ superspace chiral integral of 
a holomorphic function.
For $\CN=4$ we expect even stronger constraints, however analysis is
hampered by the lack of a $\CN=4$ superspace formalism. If such a
formalism existed,
a chiral integral would be an integral
over eight Grassman coordinates and one would expect non-renormalization
theorems for four derivative terms. 
It is nevertheless possible to work with $\CN=2$ superfields and make
additional symmetry arguments to constrain the form of the action.

This is the approach we take in this paper, where we prove a
non-renormalization theorem for terms quartic in derivatives in
theories with $\CN=4$ supersymmetry, and terms
related to these by supersymmetry. The terms of this type 
arise from an integral over all of $\CN=2$ superspace of a real
function of the superfields
\eqn\genterm{
\int d^8 \theta \CH (W, \bar W, Y, \bar Y, \tau, \tau^\dagger)~.
}
where $W$ is the field strength supermultiplet and $Y$ the adjoint hypermultiplet.
The terms involving just the $\CN=2$
vector multiplets have been much studied in previous works.
In \dewit, it was shown that the perturbative contributions 
to $\CH(W,\bar W)$ must take the form
\eqn\hterm{
\CH(W,\bar W) = \CH^0 + c (\log W^2 + g^0(W)) (\log \bar W^2 + g^0(
\bar W) )~,
}
where $\CH^0$ and $g^0$ are homogeneous and $c$ is a constant.
For $SU(2)$ $g^0=0$, and the one loop contribution to 
$\CH^0$ was determined in \dewit. The result can be understood as follows.
$\CH^0$ receives contributions only from the non-commuting off-diagonal degrees
of freedom in the vector multiplet, the W bosons. These are all massive 
if we consider maximal symmetry breaking and so 
$\CH^0$ is irrelevant if we are only interested in the massless degrees of freedom
in the maximally broken case.
The second term in \hterm\ represents everything that does not go into 
$\CH^0$. In particular, if all degrees of freedom are commuting, this term
is all there is. This second term was considered further in \dine, where it
was argued that the one-loop contributions to $\CH$ are exact
non-perturbatively for $SU(2)$. \foot{In \refs{\dorey, \diego} it was
checked that this term does not receive
non-perturbative contributions.
The constant $c$ has recently been
found to be $1/(8\pi)^2$ \refs{\periwal, \gonzalez, \butch}.}
In this work, we generalize this
result to the $SU(N)$ case and completely determine $\CH$ for the case
with maximally broken gauge symmetry.

This result is important in the context of recent studies in Matrix
theory, and in the correspondence between string theory in an anti-de
Sitter background and large $N$ $SU(N)$ $\CN=4$ supersymmetric
Yang-Mills theory \refs{\wati,\malda}. 
From the Matrix theory point of view, the
non-renormalization of these terms leads to agreement with tree-level
M-theory in
discrete
light-cone gauge further compactified on
a three-torus. The result of \dine, gives the needed
non-renormalization theorem for the $SU(2)$ case and the
complete non-renormalization theorem for general $SU(N)$ is obtained
in the present work.

Discretized light-cone quantization of
M-theory further compactified on $T^{3-n}$ is described by 
dimensionally reducing the four dimensional gauge
theory to $4-n$ dimensions. 
In this case it is not obvious that the non-renormalization
theorem will carry over because the underlying conformal invariance of 
the four dimensional theory is broken in lower dimensions.
In three dimensions, instanton effects appear, with just the right
structure predicted by the correspondence with supergravity for
$SU(2)$ \pssthree. In one dimension, a non-renormalization theorem has 
been proven by very different methods for the supersymmetric $SU(2)$ 
quantum mechanics,
where it is found that the one-loop contribution is exact \pssone. For
$SU(N)$ it has been established that the terms that appear at one-loop are
not renormalized \lowe, however the possibility
remains that other tensor structures can appear at the same order,
beyond one-loop. As we will describe below, conformal invariance and
chiral $U(1)_R$ symmetry is sufficient to rule out such extra
contributions in four dimensions.

There is a
qualitative difference between the $SU(N)$
and the $SU(2)$ case from the brane point of view. 
Consider one D-threebrane probing a generic background of $N-1$ 
others. For $N=2$ the velocity of the background D-brane may be set
to zero, giving rise to a supersymmetric background.
 It is not
surprising then that the metric induced by this brane is protected by
supersymmetry. For $N>2$ on the other hand, the background will
generically break
all supersymmetry. Nevertheless the result we find is that the metric
induced by this background is protected by supersymmetry.

\newsec{$F^4$ Terms in $\CN=4$ SUSY $SU(N)$ Gauge Theory}

In the following, the $\CN=2$ superspace formalism will be used. Each
$\CN=4$ vector multiplet consists of a $\CN=2$ vector multiplet and a
hypermultiplet in the adjoint representation. We will be interested in 
a generic point on the Coulomb branch of the theory where $SU(N)$ is
broken to $U(1)^{N-1}$ modulo Weyl transformations. The light degrees
of freedom therefore correspond to a collection of $N-1$ $\CN=4$
multiplets.

The $\CN=4$ theory is finite and conformally invariant. For a suitable 
definition of the matrix of couplings $\tau_{ij}$ and field strength 
superfields $W_i$ 
compatible with
duality, the generic gauge kinetic terms can be written as
\eqn\kinetic{
\Im \int d^2\theta d^2 \tilde \theta \tau_{ij} W_i W_j~,
}
where $\tau_{ij} = \tau C_{ij}$, where 
$\tau = {\theta \over 4\pi} +{2\pi i \over g^2}$ 
is the coupling that transforms in the standard
way under $SL(2,\IZ)$ duality and $C_{ij}$ is proportional to the
Cartan matrix. In addition there are kinetic terms for the
hypermultiplet superfields $Y_i$ and a superpotential coupling the vector and
hypermultiplet which are completely determined by supersymmetry.

Now consider terms of the form \genterm. Scale invariance implies that
$\CH$ must be dimensionless, and since the theory is $U(1)_R$ invariant
(there is no anomaly in $\CN=4$) it must transform trivially under the chiral
$U(1)_R$ symmetry. This implies strong constraints on the functions of 
the $W_i$ and $Y_i$ that can appear in $\CH$. To determine the behavior as a
function of $\tau$ we follow the argument of \seiberg, and promote
$\tau$ to a background vector superfield. Demanding scale invariance
and $U(1)_R$ invariance in this situation implies that $\CH$ cannot
depend on $\tau$ at all. We find therefore $\CH$ must be one-loop
exact.

It remains then to explicitly determine the one-loop form of
$\CH$. 
For the moment we will assume that only the scalars in the
$\CN=2$ vector multiplets have nontrivial expectation values. 
The one-loop form for $\CH(W,\bar W)$ was determined for a general
$\CN=4$ super Yang-Mills theory in terms of a momentum integral in
\gonzalez
\eqn\momentin{
\CH(W, \bar W) = \half \int {dp^2 \over {(4\pi)^2 p^2}} {\rm Tr}_A \log
\biggl( 1+ {W \bar W + \bar W W \over 2p^2} \biggr)~,
}
where the trace is in the adjoint representation which we denote by
the subscript $A$.

It is more convenient to use the fundamental representation to
describe a generic point on the Coulomb branch. 
The generators of the fundamental representation (denoted by subscript $F$)
are related to those of the adjoint representation by
\eqn\fundad{
(T_A^a)^{i \ k}_{j; \ l} = (T_F^a)^i_l \delta^k_j - (T_F^a)^k_j
\delta^i_l ~.
}
Here $a$ runs over the group generators and $i,j,k,l$ are indices in
the fundamental.
The scalar part of the  field $W$ has an expectation value
 that lives in the Cartan subalgebra
\eqn\wvev{
W^{i  \ k}_{j;  \ l} = \vev{W^a}(  (T_F^a)_i - (T_F^a)_k)
\delta^i_l \delta^k_j ~.
}
The $T_F$'s appearing here are diagonal, so are just written with a
single fundamental index.

Inserting \wvev\ into the expression that appears in the integral \momentin\
leads to
\eqn\wsqvev{
(W  \bar W + \bar W W)^{i \ k}_{j; \ l} = 2\delta^i_l \delta^k_j
(W_i -W_k)(\bar W_i - \bar W_k) ~,
}
where $W_i = W^a (T_F^a)_i$ are diagonal. Now the point is that the
right-hand side of \wsqvev\ is diagonal in adjoint indices (the trace
of \fundad\ involves the contraction of $i$ with $l$ and $j$ with $k$).
Evaluating the integral \momentin\ gives
\eqn\fans{
\CH(W,\bar W) =  
{1\over (4\pi)^2} \sum_{i<k} \log  (W_i-W_k) \log(\bar W_i - \bar W_k)~.
}
To obtain the full expression compatible with $\CN=4$ supersymmetry,
we need to generalize this expression to include
hypermultiplet couplings. This is needed to
study backgrounds corresponding to D-branes moving in
arbitrary directions. In order to do this, we first expand in $\CN=1$
superfields. It may be seen by
examining the structure of the one-loop diagrams that the correct
general expression is obtained by rotating each of the terms in the
sum \fans\ independently with respect to the manifest 
$SU(3)\times U(1)$ subgroup of the $SU(4)_R$ symmetry group of $\CN=4$ 
supersymmetry.
As expected from the supergravity point of view, the result 
corresponds to the pairwise interaction of D-threebranes. 

It should be 
noted the one-loop exactness of $\CH$
extends immediately to arbitrary gauge groups. It is then a simple matter
to carry over the calculation above to obtain $\CH$ at a generic point 
in the moduli space.

There is a singularity in $\CH$ \fans\ when $W_i \rightarrow W_k$,
which corresponds to nonabelian gauge symmetry being restored. In this 
limit, off-diagonal degrees of freedom become massless and should 
be included in the effective action. The $\CH^0$ term in \hterm\ will
in general be non-zero in this situation.

After this work was completed, we received \refs{\kulik,\buch} where
similar results are obtained for the one-loop contribution to
$\CH(W, \bar W)$.

\bigskip
{\bf Acknowledgments}

D.L. thanks the Aspen Center for Physics for hospitality while this work
was initiated.
The research of D.L. is supported in part by DOE grant DE-FE0291ER40688-Task A.

\bigskip
{\bf Note Added:}

The claim of this paper that we have determined the full perturbative
and non-perturbative form of the dimension four operator $\cal H$ is
not correct. First let us review the argument why the
non-renormalization argument works for $SU(2)$ gauge group.

Under both $U(1)_R$ and scale transformations terms in $\CH$ of the form
$$f\left(\tau,\bar{\tau}\right)\ln\left(W
\right)\ln\left(\bar{W} \right) $$ are not invariant
but get shifted. The shifted pieces are purely
holomorphic or antiholomorphic and so are killed by the $N=2$
superspace measure. However, when $\tau$ is promoted to a background
superfield this procedure does not work since the measure no longer kills
the shifted piece. This is why such a term has to be independent of
$\tau$ and thus is not renormalized. 

For groups of higher rank however, one can construct several differnt
manifestly invariant combinations of fields.
Namely, the variables $\frac{W_i - W_k}{W_j - W_l}$ are all both
$U(1)_R$ and scale invariant. This means that any function of these
variables multiplied by a function of $\tau$ will be both scale
and $U(1)_R$ invariant and hence we cannot exclude such
contributions. Even when $\tau$ is promoted to a background superfield,
such terms can be generated since they are manifestly invariant.

What we have shown is that the one-loop
contribution is given by the $\cal H$ in the text and that terms of that 
particular form 
are not renormalized neither perturbatively nor
non-perturbatively.

\listrefs
\end